\RequirePackage{snapshot}
\documentclass[10pt,twocolumn,letterpaper]{article}
\usepackage{cvpr}
\usepackage{times}
\usepackage{epsfig}
\usepackage{graphicx}
\usepackage{amsmath}
\usepackage{amssymb}

\usepackage[pagebackref=true,breaklinks=true,letterpaper=true,colorlinks,bookmarks=false]{hyperref}

 \cvprfinalcopy %

\ifcvprfinal\fi

\newcommand{\B}[1]{{\bf #1}}
\newcommand{\R}[1]{{\rm #1}}

\begin{document}

\title{ Student's T Robust Bundle Adjustment Algorithm}

\author{Aleksandr Y. Aravkin\\
University of British Columbia\\
Vancouver, BC\\
{\tt\small saravkin@eos.ubc.ca}
\and
Michael Styer\\
Stanford University\\
Stanford, CA\\
{\tt\small mstyer@cs.stanford.edu}
\and
Zachary Moratto\\
NASA Ames
Moffett Field
{\tt\small zachary.m.moratto@nasa.gov}
\and
Ara Nefian\\
Carnegie Mellon University and NASA Ames
Moffett Field, CA
{\tt\small ara.nefian@nasa.gov}
\and
Michael Broxton\\
Carnegie Mellon University and NASA Ames
Moffett Field, CA
{\tt\small michael.broxton@nasa.gov}
}
\maketitle
\begin{abstract}
Bundle adjustment (BA) is the problem of refining a visual
reconstruction to produce better structure and viewing parameter
estimates.  This problem is often formulated as a nonlinear least
squares problem, where data arises from interest point matching.
Mismatched interest points cause serious problems in this approach, as
a single mismatch will affect the entire reconstruction. In this
paper, we propose a novel robust Student's t BA algorithm (RST-BA).
We model reprojection errors using the heavy tailed Student's
t-distribution, and use an implicit trust region method to compute the
maximum {\it a posteriori } (MAP) estimate of the camera and viewing
parameters in this model.  The resulting algorithm exploits the sparse
structure essential for reconstructing multi-image scenarios, has
the same time complexity as standard $L_2$ bundle adjustment ($L_2$-BA), 
and can be implemented with minimal changes to the standard least squares 
framework. We show that the RST-BA is more accurate than
either $L_2$-BA or $L_2$-BA with a $\sigma$-edit rule for 
outlier removal for a range of simulated 
error generation scenarios. The new method has
also been used to reconstruct lunar topography using data
from the NASA Apollo 15 orbiter, and we present visual and
quantitative comparisons of RST-BA and $L_2$-BA methods on this
application. In particular, using the RST-BA algorithm we were
able to reconstruct a DEM from unprocessed data with many outliers
and no ground control points, 
which was not possible with the $L_2$-BA method. 
\end{abstract}

\section{Introduction}

\begin{figure*}[t]
\begin{center}
{\includegraphics[scale=0.5]{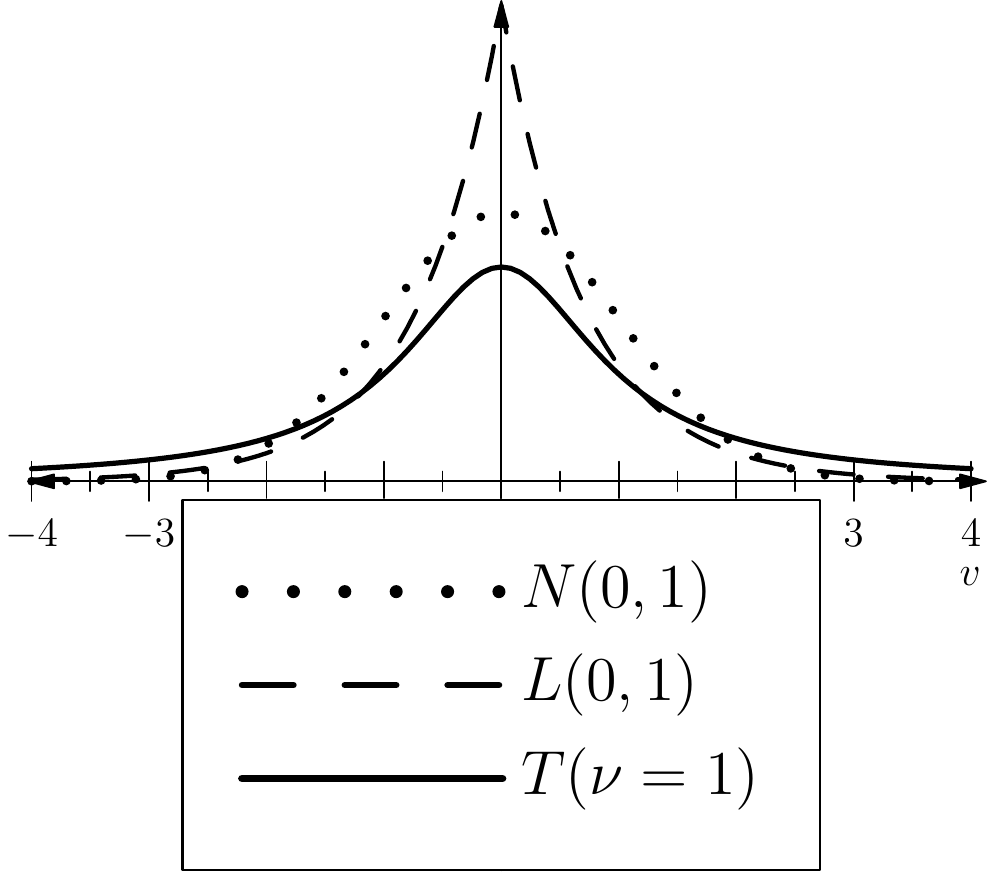}}
{\includegraphics[scale=0.5]{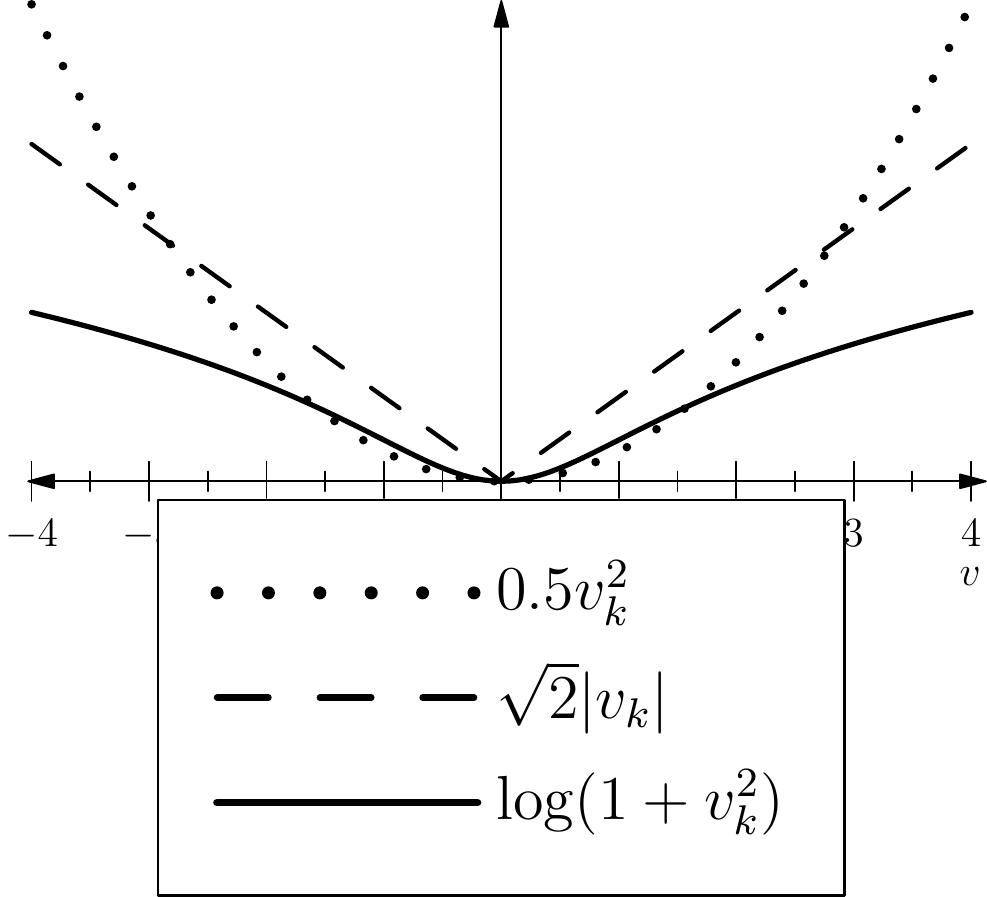}}
{\includegraphics[scale=0.5]{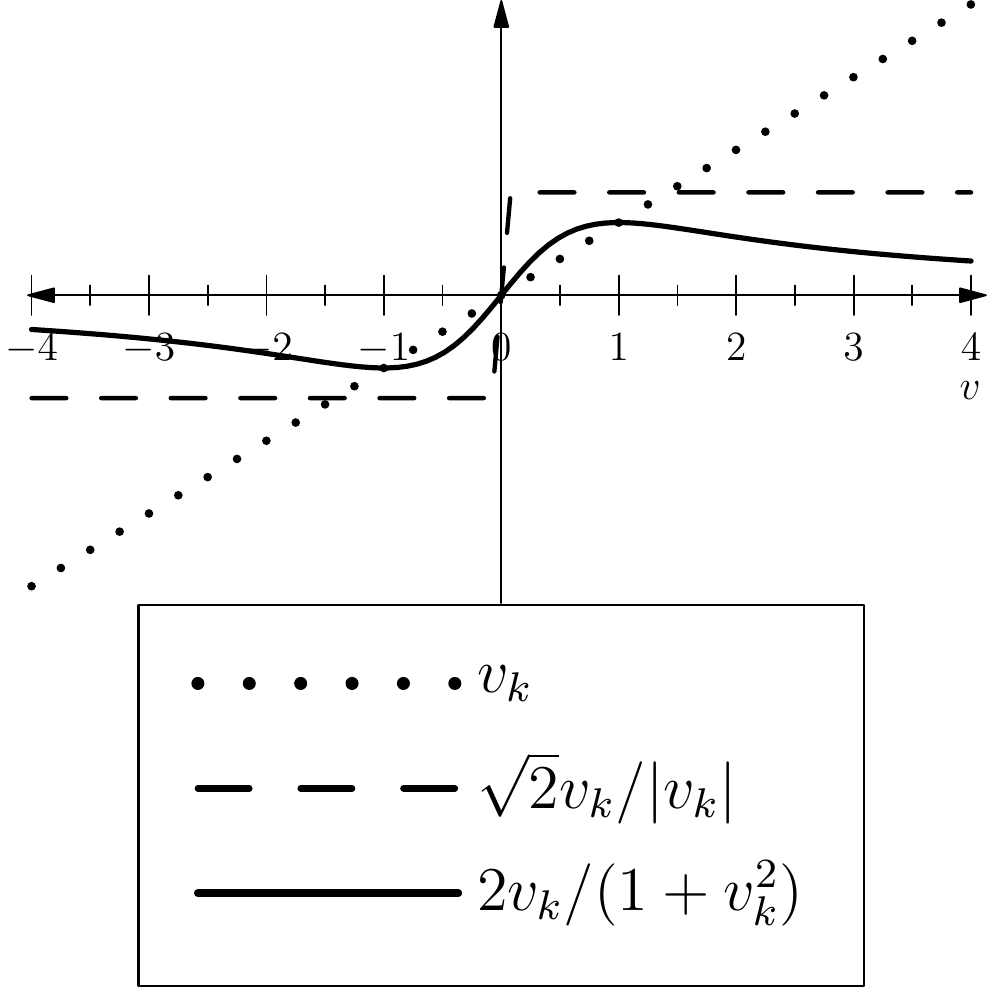}}
\caption{\label{GLT}
Gaussian, Laplace, and Student's t Densities, Corresponding Negative Log
Likelihoods, and Influence Functions (for scalar $v_k$)
}
\end{center}
\end{figure*}

Bundle adjustment is a large sparse geometric parameter estimation problem, in
which parameters are 3D feature coordinates and camera poses.  
Classically, bundle adjustment is formulated in the nonlinear least squares
framework (see \cite{Triggs00} and sources cited within).  The goal of bundle
adjustment is to refine a visual reconstruction by identifying a sparse cloud
of tie-point features in multiple images, matching tie-points common to
several images, and adjusting camera and world point parameters simultaneously
to improve the reconstruction.  Prior information about camera parameters and ground control points are incorporated using the Bayesian modeling framework,
and camera parameters and 3D world coordinates of the tie-point features are estimated using the matched tie-point features as data.

The motivating problem for our work in robust bundle adjustment is to
produce a robust reconstruction in the presence of outliers generated
due to the misidentification of tie-points across images. It is
virtually impossible to ensure that automated algorithms for finding
and matching tie-point features always generate correct 
correspondences.  A single wrong identification can lead to a large
phantom error which dominates other `good' data in the least squares
framework.

To derive our approach, we begin with a 
statistical model for the reprojection errors in pixel space as well as
initial camera parameter errors and ground control point errors, and then find
a maximum {\it a posteriori} estimate for this model using an implicit trust
region optimization method.  We model both reprojection errors and prior
uncertainty on cameras and ground control points as distributed according to
the Student's t-distribution. 

There are many approaches that try to incorporate the idea of `robust cost' or
`fat tails' into the bundle adjustment framework (see \cite{Triggs00},
\cite{Hartley2004}). Most of these schemes require some iterative
re-weighting of the least squares objective.  However, these approaches
lack a progress metric or any optimality guarantees, which limits possibilities
for algorithm design and convergence criteria. 
We begin instead with distributional assumptions on reprojection 
errors, derive a corresponding {\it maximum a posteriori} (MAP) optimization
problem, and develop an algorithm to solve it. 
Our approach takes advantage 
of the sparse structure described in ~\cite{Lourakis09:sba}
and ~\cite{Hartley00:bundleadjustment}, and determines the best camera
parameters by solving a maximum likelihood (ML) problem.
using an algorithm that
still follows the main idea in \cite{Madsen} 
while exploiting the sparse
structure in \cite{Hartley2004}. 

The resulting robust Student's t BA (RST-BA) preserves the sparse
structure and efficiency of the method used in~\cite{Lourakis09:sba}, 
allowing the robust 
implementation to run in a comparable time to the standard
BA method.  Compared with automated outlier removal schemes and ad-hoc
reweighting using `robust-cost' functions, RST-BA is faster and more
accurate both on synthetic data and in real applications. Moreover,  
it is straightforward to implement with minimal changes to the 
standard least-squares framework.

Mismatches in tie-point features is a well known problem, and there are a
variety of approaches in the literature for robustification and dealing with
outliers. Threshold-based outlier removal is a
common approach with many variants (\cite{Sunderhauf05}, \cite{Mayer03},
\cite{Mayer05}, \cite{Mayer06}, \cite{Cumani06}, \cite{Fukaya08},
\cite{Lee09}, \cite{Ke05}).  After one or more iterations of bundle
adjustment, observations with residuals exceeding some predetermined threshold
are removed from the dataset and the bundle adjustment optimization is run
again on the remaining observations.  The choice of threshold depends on the
particular problem and dataset.  \cite{Sunderhauf05}, for instance, uses a
threshold of 1.5 standard deviations from the mean residual error, while
\cite{Mayer06} discards residuals with size to variance ratio greater than 3.
Generally, some $\sigma$-edit rule is applied, with a typical range between 1
to 2 standard deviations.  However, there are problems with using
$\sigma$-edit rules even in the linear regression context \cite[Chapter
1]{Mar}, and it is particularly limiting in nonlinear problems because
outliers affect the initial fit, which is then used for classification of outliers and inliers.

Another common approach is iterative reweighting (\cite{Fua99}, \cite{Fua00},
\cite{Mayer05}, \cite{Mouragnon09}).  In this method, the observations are
weighted at each iteration in inverse proportion to their residual errors.
Observations may also be removed according to a $\sigma$-edit rule with a
higher threshold \cite{Mayer05}.  Closely related to this approach is that of
`robust cost functions', which is discussed in \cite{Hartley2004} and
implemented in \cite{Brandt06} and \cite{Cumani06}. 
The main problems with ad hoc reweighting according to 
such cost functions is the lack of a convergence
theory that guarantees the algorithm will stop at a stationary 
point of the objective, and in many cases 
the lack of an explicit objective function.

Model-based bundle adjustment is yet another robust technique, in which prior
information about the surface under reconstruction is used to aid in the
adjustment \cite{Fua00}.  \cite{Zhang06} proposes a bundle adjustment
method which does not involve solving for the camera parameters, in order to
reduce numerical instability and decrease sensitivity to noise. We do not
consider these approaches, since we have no prior information on the surface we are
reconstructing, and refinement of the camera parameters is essential
to our application. 

The paper proceeds as follows. In Section \ref{MathBA} we formulate
standard bundle adjustment as a nonlinear least squares problem that can 
be solved with $L_2$-BA.  In Section \ref{StudentT} we describe the 
multivariate Student's t-distribution. In Section \ref{MLE}, we formulate
robust bundle adjustment as the maximum {\it a posteriori}
likelihood problem and develop the RST-BA algorithm to solve it. 
In Section \ref{Numerics}, we use simulated data to
compare RST-BA with the standard $L_2$-BA algorithm, and 
with $L_2$-BA combined with a $2$-sigma edit rule for
removing outliers.  
In Section \ref{RealData}, we show the results of applying the
RST-BA algorithm to the problem of reconstructing lunar topography from NASA
Apollo 15 orbital imagery, and compare the results with results obtained from
$L_2$-BA for both processed and unprocessed data.

\section{Mathematical Model for BA}
\label{MathBA}

Suppose that we have $m$ images taken by a moving camera, 
or equivalently by $m$ cameras with different poses (locations
and attitudes). 
Suppose that multiple tie-point features are identified in each image,  
and that a single feature may appear in several images. 
Denote by $S$ the control network of indices describing 
the image tie-point matching, so that $(i,j) \in S$ if 
feature $i$ was seen in image $j$. Let $x_j$ denote the 
pose of the $j$th camera, and 
let $y_i$ denote the 3D world coordiantes of feature $i$.
Let $z_{ij}$ be the pixel coordinates of the projection
$h(x_j, y_i)$ of feature $i$ onto image $j$, 
$\epsilon_{ij}$ be the reprojection error 
$z_{ij} - h(x_j, y_i)$, and $\Sigma_{ij}$
be the covariance matrix associated to $\epsilon_{ij}$. 
Define $x_j^0$ and $y_i^0$ to be prior estimates
of camera parameters and ground control points, with 
covariances $\Omega_j$ and $\Phi_i$, 
respectively. All the data except for $\{x_j\}$ and $\{y_i\}$ are known and given. 
The standard statistical model for the BA problem is as follows:
\begin{equation}
\label{BAModel}
\begin{array}{lll}
z_{ij} &=& h(x_j, y_i) + \epsilon_{ij}\\
\vspace{.1cm}
\epsilon_{ij} &\sim& (0, \Sigma_{ij})\\
\vspace{.1cm}
x_j &\sim &(x_j^0, \Omega_j)\\ 
\vspace{.1cm}
y_i &\sim &(y_i^0, \Phi_i)
\end{array}
\end{equation}
The standard $L_2$ approach to bundle adjustment is to 
assume that $\epsilon_{ij}$, $x_j$, and $y_i$
are all normally distributed, and then 
the maximum {\it a posteriori}
likelihood solution is equivalent 
to minimizing the objective 
\begin{eqnarray}
\label{NLSobjective}
\mbox{min} &\;\;&
\nonumber	\frac{1}{2}\sum_{(i,j) \in S}
[z_{ij} - h(x_j, y_i)]^T
\Sigma^{-1}_{ij}
[z_{ij} - h(x_j, y_i)]\\
       &&\;\;\;\;+ 
\sum_{j}[x_j^0-x_j]^T
\Omega^{-1}_j
[x_j^0 - x_j]\\
\nonumber &&\;\;\;\;+
\sum_{i}[y_i^0-y_i]^T
\Phi^{-1}_i
[y_i^0 - y_i]
\\
\nonumber \; \mbox{w.r.t.} &\;\;& \{x_j\}, \{y_i\}.
\end{eqnarray}

The prior terms for camera parameters and ground control 
points are added to deal with gauge freedom.  
Ground control points 
can be used when 3D coordinates of certain tie-point features are 
well known, which is the case for Lunar topography data. 
Note that if there is no prior information on 
a particular $x_j$ or $y_i$, we simply set the 
corresponding  $\Omega_j^{-1}$ or $\Phi_i^{-1}$ to $\B{0}$
in (\ref{NLSobjective}).  

The standard approach to bundle adjustment is to minimize the objective (\ref{NLSobjective}) using implicit trust region methods, 
and in particular variants of
the Levenberg-Marquardt method are very popular (see \cite{Madsen}, \cite{Hartley2004}, \cite{WiSeb}, \cite{Noce}, \cite{Flet} 
for more details on these methods). 
For our implementation of $L_2$-BA 
we use a particular variant of the Levenberg-Marquard detailed in 
\cite[Algorithm 3.16]{Madsen}, which is also used in the 
SBA implementation~\cite{Lourakis09:sba}.  
The method of choosing a cloud of points that `links'
the images together gives rise to a sparse structure, 
and we exploit this structure as described in 
\cite[Algorithm A6.4]{Hartley2004}.

\section{Student's t Approach}
\label{StudentT}

We introduce the following notation: for a vector 
\(u \in \B{R}^n\)
and any positive definite matrix 
\(M \in \B{R}^{n\times n}\), let
\(\|u\|_M := \sqrt{u^TMu}\).
We use the following generalization of the Student's 
t-distribution: 

\begin{eqnarray}
\label{StudentDensity}
\B{p}(\epsilon | \mu)
&=&
 \frac{\Gamma (\frac{s + m}{2})}
{\Gamma(\frac{s}{2})\det[\pi s R]^{1/2}}
\left(
1 + \frac{\|\epsilon - \mu \|_{R^{-1}}^2}{s}
\right)^{\frac{-(s + m)}{2}}
\end{eqnarray}
where $\mu$ is the mean parameter, $s$ is the degrees of freedom,
$m$ is the dimesion of the vector $\epsilon$, 
$R$ is a positive definite matrix, and 
$\sqrt{R}$ or $R^{1/2}$ denotes a Choleski factor; 
i.e., $\sqrt{R} \sqrt{R}^\R{T} = R^{1/2} R^{\R{T}/2} = R$.
A comparison of this distribution 
with the Gaussian distribution assumed in (\ref{NLSobjective}) and 
the Laplace distribution appears in Figure \ref{GLT}. 
Note that the Student's t-distribution
has much thicker tails than the others,  
and that its influence
function is redescending (see \cite{Mar} for a discussion of 
influence functions). 

The main idea of the RST-BA algorithm is to assume that 
reprojection errors $\epsilon_{ij}$ come from 
the Student's t-distribution (\ref{StudentDensity}). 
We also assume the Student's t-distribution prior 
for the initial camera parameters $\{x_j^0\}$ and 
ground control points $\{y_i^0\}$.  
The intuition behind this approach is that 
extreme observations are much more likely 
in the Student's t model than in the Gaussian model. 
Therefore, a large residual will affect the overall fit less
if fitting is done in model (\ref{StudentDensity}). 
See~\cite{AravkinThesis2010} for more details.

\section{Maximum Likelihood Formulation}
\label{MLE}

\begin{table*}
\caption{
\label{SimulationResults}
Relative mean $\mu$ and standard deviation $\sigma$ 
of MSE calculated over 1000 runs for $L_2$-BA, $L_2$-BA
with the 2$\sigma$-edit rule (2$\sigma$-BA), 
and RST-BA, presented as: $\mu$ ($\sigma$). 
Error values for world points and camera
XYZ parameters respectively are presented relative to the 
error incurred by $L_2$-BA in the nominal case, shown 
in bold, i.e.
where reprojection errors added were distributed as $N(0,1)$.
}
\begin{center}
\begin{tabular}{|c|c|c|c|c|c|c|}\hline
Noise Type 
& \multicolumn{3}{|c|} {World Points} 
& \multicolumn{3}{|c|} {Camera XYZ}
\\ \hline
& $L_2$-BA 
& $2\sigma$-BA
& RST-BA 
& $L_2$-BA
& $2\sigma$-BA
& RST-BA
\\\hline
$N(0,1)$  
& {\bf 1.0 (1.3)}
& 1.0 (1.2)
& 1.0 (1.3)
& {\bf 1.0 (0.9)}
& 0.8 (0.7)
& 0.7 (1.4)
\\\hline
$.95 N(0,1) + .05 N(0,4)$
& 1.3 (1.6) 
& 1.2 (1.5)
& 1.1 (1.4)
& 6.3 (7.6)
& 2.7 (4.1)
& 3.5 (13.3)
\\\hline
$.9 N(0,1) + .1 N(0,4)$
& 1.5 (1.7)
& 1.5 (1.9) 
& 1.4 (1.7) 
& 11.5 (12.5)
& 5.6 (7.6)
& 5.9 (18.1)
\\\hline
$.95 N(0,1) + .05 N(0,10)$
& 2.7 (3.4)
& 1.8 (2.0) 
& 1.2 (1.4) 
& 69 (62)
& 23 (27)
& 7.3 (23)
\\\hline
$.9 N(0,1) + .1 N(0,10)$
& 3.6 (4.6)
& 2.7 (3.0)
& 1.4 (1.6)
& 101 (76)
& 49 (42)
& 16.5 (34)
\\\hline
$.95 N(0,1) + .05 N(0,50)$
& 39 (45)
& 21 (30)
& 1.9 (1.7)
& 580 (380) 
& 306 (242)
& 12 (23) 
\\\hline
$.9 N(0,1) + .1 N(0,50)$
& 60 (63)
& 44 (47)
& 2.5 (2.1)
& 740 (510)
& 470 (300)
& 20 (36)
\\\hline
Student's t, df = 4
& 12.3 (13.5)
& 12.2 (15.1)
& 8.9 (10.2)
& 240 (150)
& 190 (130)
& 38 (60)
\\\hline
\end{tabular}
\end{center}
\end{table*}

Maximizing the likelihood for our model (\ref{BAModel}) 
is equivalent to minimizing the associated negative log likelihood 
\[
-\log \B{p}(\{\epsilon_{ij}\}) 
-\log \B{p}(\{x_j^0 - x_j\})
-\log \B{p}(\{y_i^0 - y_i\})
\]
Dropping the terms that do not depend on \(\{x_j\}\) or
 \(\{y_i\}\) our objective is 
\begin{equation}
\label{fullObjective}
\begin{array}{lll}
&\frac{1}{2}\sum_{(i,j) \in S}(s_{ij} + 2)
\log\left[
1 + \frac{1}{s_{ij}}\|z_{ij} -h(x_j, y_i)\|_{\Sigma_{ij}^{-1}}^2
\right]\\
&+ \frac{1}{2}\sum_{j}(r_{j} + 6)
\log\left[
1 + \frac{1}{r_j}\|x_j^0 - x_j\|_{\Omega_j^{-1}}^2
\right]\\
&+ \frac{1}{2}\sum_{i}(q_{i} + 3)
\log\left[
1 + \frac{1}{q_i}\|y_i^0 - y_i\|_{\Phi_i^{-1}}^2
\right]

\end{array}
\end{equation}
where $s_{ij}$, $r_j$, and $q_i$ are known degrees of freedom
parameters  in model (\ref{StudentDensity}) associated 
to observations $z_{ij}$, 
prior camera parameters $x_j^0$, and ground control 
points $y_i^0$, respectively. 
The constants $2$, $6$ and $3$ that appear in (\ref{fullObjective})
are the dimensions of the pixel coordinates,  
camera poses, and world points, respectively.  

Minimizing objective  (\ref{fullObjective}) provides 
maximum  {\it a posteriori} (MAP) likelihood estimates  
for parameter vectors $\{x_j\}$ and $\{y_i\}$
in the Student's t model (\ref{StudentDensity}). 

Now we describe an implicit trust region algorithm
for minimizing (\ref{fullObjective}). 
Given a sequence of column vectors $\{ u_k \}$
and matrices $ \{ T_k \}$ we use the notation
\[
\R{vec} ( \{ u_k \} )
=
\begin{bmatrix}
u_1 \\ u_2  \\ \vdots \\ u_N
\end{bmatrix}
\; , \;
\R{diag} ( \{ T_k \} )
=
\begin{bmatrix}
T_1    & 0      & \cdots & 0 \\
0      & T_2    & \ddots & \vdots \\
\vdots & \ddots & \ddots & 0 \\
0      & \cdots & 0      & T_N
\end{bmatrix}
\]
We define 
\[
\begin{array}{rcl}
c              &=& \R{vec}(\{x_j\}, \{y_i\}). \\
\end{array}
\]
We will now refer to objective (\ref{fullObjective})
as $F(c)$. In order to minimize (\ref{fullObjective}),
we implement an iterative method of the form 
\begin{equation}
\label{iterScheme}
\begin{array}{lll}
\hat c
&=& 
c^k -
\left(H^{k}\right)^{-1}
\nabla F(c) \\ 
c^{k+1} 
&=& 
\begin{cases}
\hat c &
\text{if } F(\hat c) < F(c^k)\\
c^k &
\text{otherwise.}
\end{cases}
\end{array}
\end{equation}
where $k$ indexes the iterations, and 
$H^k$ is a particular positive definite matrix
described below,  
which one may think of as a Hessian approximation
to $\nabla^2 F(c^k)$.   

Let 
$J^k 
= 
\begin{bmatrix} A^k & B^k\end{bmatrix}$, 
where 
$A_{ij}^k 
= 
\partial_{x_j}h(x_j^k, y_i^k)$ and 
$B_{ij}^k 
= 
\partial_{y_i}h(x_j^k, y_i^k)$. 
Define weights 
\begin{equation}
\label{weights}
\begin{array}{lll}
g_i^k = \frac{q_{i} + 3}
{q_{i} + \left\|y_{i}^{0} - y_{i}^k\right\|_{\Phi_{j}^{-1}}}
&\;,\;&
\varrho_{j}^k 
= 
\frac{r_{j} + 6}
{r_{j} + \left\|x_{j}^{0} - x_{j}^k\right\|_{\Omega_{j}^{-1}}}\\\\
\rho_{ij}^k 
= 
\sqrt{
\frac{s_{ij} + 2}{s_{ij} + \left\|\epsilon_{ij}^k
\right\|_{\Sigma_{ij}^{-1}}}
}

\end{array}
\end{equation}
Let $\tilde A_{ij}^k = \rho_{ij}^kA_{ij}^k$ and 
$\tilde B_{ij}^k = \rho_{ij}^kB_{ij}^k$. Let 
$\tilde A^k$ and $\tilde B^k$ be 
matrices with block components 
$\tilde A_{ij}^k$
and $\tilde B_{ij}^k$, 
respectively, and define
\(\tilde J^k = 
\begin{bmatrix} 
\tilde A^k & \tilde B^k
\end{bmatrix}\). 
Define
\begin{equation}
\label{HessianApprox}
\begin{array}{lll}
H^k 
&=& 
\left(\tilde J^k\right)^T 
\Sigma^{-1} 
\left(\tilde J^k\right) 
+
\R{diag}\left(
\{\varrho^k_{j}\Omega_j\}
\right)\\
&+&
 \R{diag}
\left(
\{g^k_{i}\Phi_i\}
\right)
+
\lambda^k I
\end{array}
\end{equation}
where $\lambda^k$ is a regularization parameter similar to the 
Levenberg-Marquardt method, and is updated according to the rule defined 
in Algorithm 3.16 of \cite{Madsen}. Specifically, $\lambda$
increases quickly when the iteration (\ref{iterScheme}) fails to 
improve the objective function (\ref{fullObjective}), and 
otherwise is adjusted according to the rule 
\begin{equation}
\label{lambdaUpdate}
\lambda^{k+1} 
= 
\lambda^k \max(\frac{1}{3}, 1-(2 \phi^k -1)^3)
\end{equation}
where $\phi^k$ is the ratio of improvement predicted by 
the quadratic model with Hessian $H^k$ to the actual 
improvement $F^{k} - F^{k+1}$. 

From this presentation, it is clear that the RST-BA algorithm can be
implemented by a simple reweighting of the data structures already present in
$L_2$-BA, and so RST-BA takes about the same time per
iteration as $L_2$-BA.  
The algorithm terminates when all the components of $\nabla
F$ are below $10^{-6}$.  In practice, a hard iteration limit is set, since the
problems are large and it is rarely necessary to solve them exactly.  We
followed this approach in testing and simulation. 

\section{Numerical Simulations}
\label{Numerics}

The RST-BA code used for the simulated and real tests 
is currently implemented as part of 
Nasa VisionWorkbench~\cite{VisionWorkbench}.
Since our target application is the reconstruction of the lunar surface 
from Apollo orbital imager data, 
our synthetic data was modeled in a similar context. 
Camera positions were generated in a sequence incremented
along the camera $x$-axis, with the $z$-axis of the
camera coordinate system defined to point toward the lunar surface. 
The $x$-increment was calculated to yield the desired overlap 
between camera fields of view, 
to guarantee that each point on the surface was seen by at least two
cameras. Given specifications for the camera elevation and
location, a synthetic surface region was calculated, 
bounded by the camera field of view in the $y$-direction, 
the combined fields of view of the second through the penultimate cameras 
in the $x$-direction, and an estimate
of minimum and maximum lunar surface height in the $z$-direction.
$3$-dimensional world points were then randomly generated within the 
volume bounded by this surface. 
Finally, each generated $3$-dimensional point was projected into the image
plane of each camera in which it was visible, 
giving a set of image coordinates
for each point and camera pair.

After generation of the synthetic 3D world points, we added several
kinds of noise to the ``observations'' made in the simulated system.
In the lunar surface reconstruction context, observations include the
image coordinates of each visible point, and extrinsic camera
parameters that are known up with some precision from the Apollo mission
telemetry.  Our data generator perturbs image coordinates, camera
positions, and camera pose according to nominal Gaussian distributions
with specified variance in order to simulate measurement noise and
camera uncertainty.  To test the robustness of our algorithms against
mistakes in the data, we also introduced outliers in the simulated
errors according to error schemes we describe below.
\begin{enumerate}
\item
Nominal conditions:
The reprojection errors were generated using the normal distribution
$\epsilon_{ij} \sim \B{N} ( 0 , 1 )$.

\item
Contaminated normal:
The reprojection errors were generated using 
a mixture of two normals, i.e., 
\[
\epsilon_{ij} \sim (1 - p ) \B{N} (0, 0.25 ) + p \B{N} ( 0 , \phi )
\]
for values of $p \in \{ 0.05 , 0.1 \}$
and values of $\phi \in \{ 4, 10, 50 \}$.

\item Student's t-distribution:
The reprojection errors were generated using Student's t-distribution
with \( df = 4 \).
\end{enumerate}
For each run of each experiment, we ran the $L_2$-BA algorithm as the
baseline, along with $L_2$-BA combined with a
$2\sigma$-edit rule (removing `outliers' that were two standards of deviation
away from the mean and refitting), and the RST-BA algorithm.  All degrees of
freedom parameters for RST-BA were set at $4$ for all of the experiments.  

The results for our simulated fitting are presented in
Table~\ref{SimulationResults}.
Each experiment was performed 1000 times, and we provide the
relative median Mean Squared Error (MSE) value and standard deviation
for the difference between `ground truth' and the final 
estimates of the algorithms, for the 3D world points data
and for the camera location $(x,y,z)$ data. We left the 
camera pose parameters fixed at their true values 
during the experiment, 
by placing a very strong prior on them. 
The relative MSE for the world points is defined by 
\begin{equation}
	\left(\frac{1}{N} \sum_{k=1}^N
		\|X_k  - \hat{X}\|_2^2\right)/\left(\B{MSE_0}\right)
\end{equation}
where $N$ is the total number of 3D world points,
 \(X_k\) is the $k$-th `true' world point,  
\( \hat{X}_k\) is the estimate, 
and $\B{MSE_0}$ is the MSE of the baseline
BA method in nominal conditions. 
The relative MSE measure for camera
coordinates is similarly defined. 

The $L_2$-BA method with the $2\sigma$-edit
rule works as well or better than $L_2$-BA alone. 
When the variance of the outliers is very large, 
the $2\sigma$-edit rule cuts the relative error nearly 
in half, for both world points and camera parameters. 
The RST-BA algorithm works about as well as the $2\sigma$-edit
rule for cases with small outliers, but as the variance
of the outliers grows, RST-BA cuts the relative 
error by a factor of 30 -- an order
of magnitude improvement over the $2\sigma$-edit rule. 
When the errors are actually generated from the Student's 
t-distribution, RST-BA cuts the camera 
error by a factor of 6 relative to the standard, 
and achieves a small improvement for the world points.

\begin{figure}[t]
\begin{center}
{\includegraphics[scale=0.25]{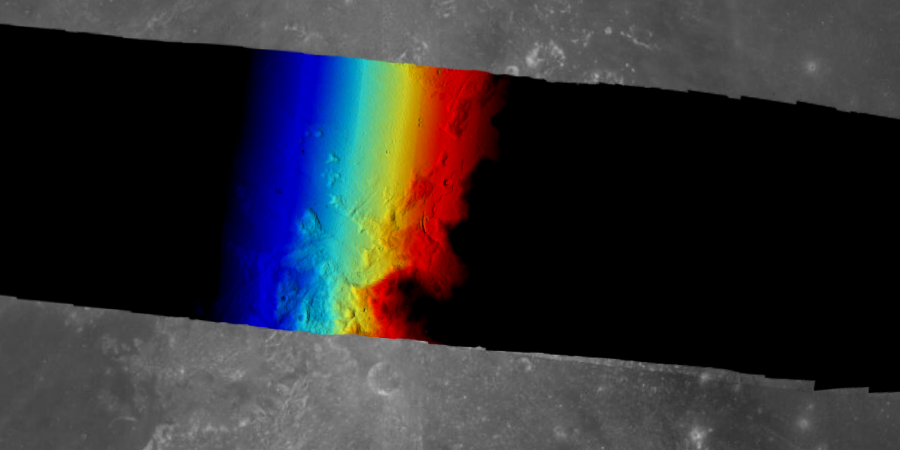}}
(a)
{\includegraphics[scale=0.25]{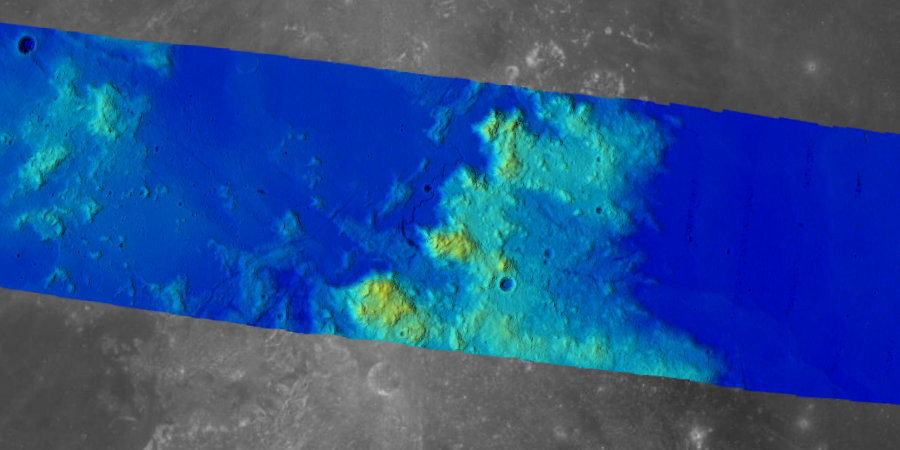}}
(b)
{\includegraphics[scale=0.25]{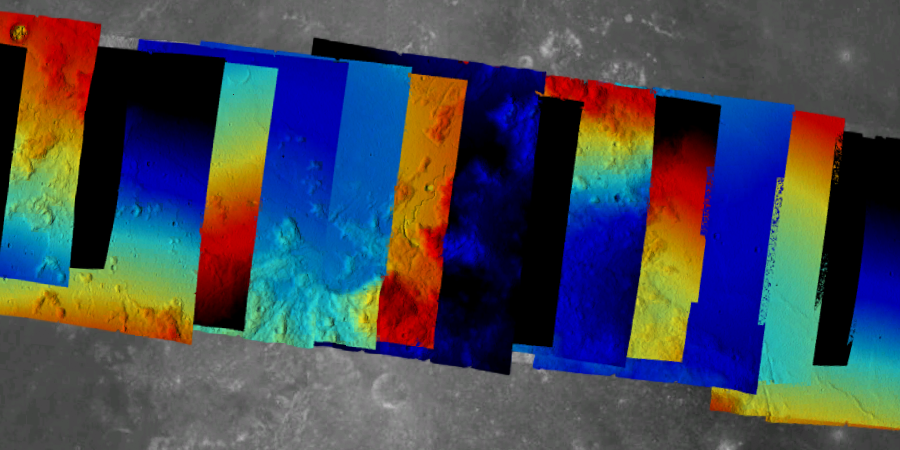}}
(c)
{\includegraphics[scale=0.25]{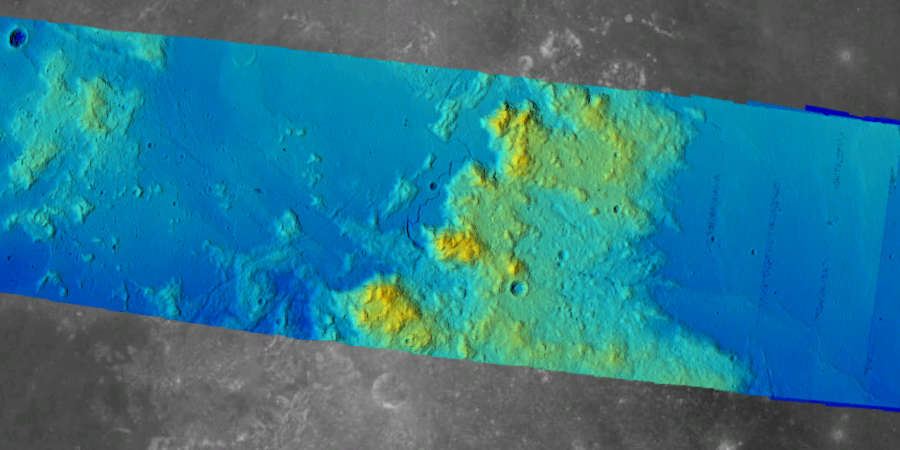}}
(d)
\caption{\label{realFigs} Surface reconstruction from Orbit 33 images. From
top to bottom: (a) $L_2$-BA, processed data set;  (b) RST-BA,
processed data set; (c) $L_2$-BA, unprocessed data set; (d) RST-BA,
unprocessed data set. Red indicates high elevation, blue indicates low
elevation. Black indicates elevations out of the range of the color map.
Ground control points were not used in these experiments.
  }
\end{center}
\end{figure}

\section{Bundle Adjustment in Orbital Imagery}
\label{RealData}

To check the performance of the RST-BA algorithm on real data, we used
imagery captured by the Apollo Metric Camera (AMC) on board the NASA Apollo 15
orbiter. The AMC is a frame camera with a 74 degree field of view that
captured snapshots of the Moon's surface at regular intervals. This resulted
in overlap between images of 80\%.  The Apollo-era satellite tracking network
was highly inaccurate by today's standards, with errors estimated to be
2.04-km for satellite station positions and 0.002 degrees for pose estimates.
This creates a need for refinement via bundle adjustment in order to create
consistent 3D models between stereo pairs (see Figure \ref{realFigs}).  The
specific frames processed were AS15-M-1089 through AS15-M-1159, which were
part of Apollo 15's 33rd orbital revolution\cite{Lawrence08}.

In order to test the effectiveness of $L_2$-BA against RST-BA, two datasets of image measurements were created:

\begin{enumerate} 
\item
{\em Processed Apollo tie-point data} was created with extensive
processing and cleaning. First, tie points were automatically detected
with the SURF \cite{Bay08} algorithm. Then outliers were removed using
the RANSAC algorithm \cite{Fischler81}. Finally the tie points
were thinned down to 500 matches between pairs by removing the weakest
matches while ensuring that the tie points remained evenly
distributed across each image.  
\item
{\em Unprocessed Apollo tie-point data} was created using 
the Interest Point
detection algorithm based on SIFT \cite{Lowe04}.  No outlier rejection
was done in this case, yielding a data set with up to 50\% outliers.
\end{enumerate}

$L_2$-BA and RST-BA were tested with both
processed and unprocessed data sets. Results of these tests are shown in
Table~\ref{triError}. Here, triangulation error is a measure of the average
distance between the closest point of intersection of two forward projected
rays for a set of tie-points. A decrease in triangulation error indicates a
substantial improvement in the self-consistency of the DEMs in the data set.

After bundle adjustment was complete, we processed the imagery using 
stereo reconstruction tools \cite{Nefian2009}
to produce dense topography of the lunar surface.
This 3D reconstruction used the improved camera extrinsic parameters from
bundle adjustment to produce a more consistent, seamless mosaic of 3D
topographic models.  Figure~\ref{realFigs} shows these results with various
bundle adjusment tests.  Topography is represented by a color-map with red
indicating high elevation and blue low elevation.

The original (unadjusted) camera parameters show clear discontinuities between
adjacent models that are due to the uncertainties in the original Apollo
tracking data. While {\em Processed} data yields reasonable results when
ground control points are used, we did not use these points 
to emphasize that RST-BA can be used in their absence while   
$L_2$-BA cannot. Without ground control points, $L_2$-BA
found a `kink' in the DEM, which is responsible for the 
black sections visible in Figure \ref{realFigs} while 
RST-BA was able to reconstruct the DEM.

The results from the {\em unprocessed} data set, which contained nearly 50\%
outliers, show a stark difference between the two approaches.  The 
$L_2$-BA algorithm failed to create any improvement.  Instead, the
outliers caused severe and unpredictable distorition of the camera parameters.
The RST-BA algorithm, on the other hand, was nearly unaffected, and produced
results remarkably similar to those produced from the {\em processed} data
set.  Table~\ref{triError} shows that median triangulation error were only
slightly higher for the {\em unprocessed} data than they were for the 
{\em processed} data.  This data suggests that RST-BA is significantly more
robust to outliers than the standard bundle adjustment technique.

\section{Conclusion}

We have proposed RST-BA, a robust bundle adjustment algorithm, based on the
Student's t distribution, for performing bundle adjustment in
the presence of outliers in tie-point matching. RST-BA preserves the sparse
structure, and hence the speed, of $L_2$-BA, and can be
implemented by simple modifications to the $L_2$-BA
algorithm. Our test results on both synthetic 
and real data show that when the
data have been preprocessed to remove 
outliers in the tie-point matches,
RST-BA outperforms $L_2$-BA by a small margin, and on
unpreprocessed data with numerous outliers, RST-BA outperforms 
$L_2$-BA by a significant margin. 
RST-BA demonstrates significant advantages in both speed and accuracy
of results over both $L_2$-BA and $L_2$-BA with a $2\sigma$-edit rule, 
and can be used to reconstruct DEMs without data preprocessing 
and without ground control points. 
In future work, we will perform an extended comparison of RST-BA with ``robust'' methods such as Cauchy re-weighting in addition
to the $2\sigma$-edit rule. We will also work on the problem
of estimating the degrees of freedom parameters, which are currently
assumed to be known by the RST-BA algorithm. 

\begin{table*}
\begin{center}

\begin{tabular}{|c|c|c|c|c|c|c|c|}\hline
Dataset
&Algorithm
& \multicolumn{3}{|c|} {Start Triangulation Error} 
& \multicolumn{3}{|c|} {End Triangulation Error}
\\ \hline

&
& Min
& Median
& Max
& Min
& Median
& Max
\\ \hline
Processed 
&$L_2$-BA
& 0.0616
& 714.18
& 134234
& 0.0008
& 36.309
& 178636
\\ 
 
&RST-BA
& 0.0616
& 714.18
& 134234
& 0.0002
& {\bf 14.329}
& 132966
\\ \hline
Unprocessed 
& $L_2$-BA
& 0
& 729.87
& 239452
& 0
& 4409.9
& 242179
\\ 
 
& RST-BA
& 0
& 729.87
& 239452
& 0
& {\bf 14.404}
& 240647
\\ \hline

\end{tabular}
\end{center}
\caption{Triangulation errors for the Apollo DEM 
reconstruction, using {\it processed} and {\it unprocessed} 
Apollo tie-point data. RST-BA performs much better 
than $L_2$-BA for {\it unprocessed} data, and better
than $L_2$-BA for {\it processed} data. }
\label{triError}
\end{table*}

\section{Acknowledgements}

We would like to thank our colleagues at the Arizona State University
for supplying high resolution scans of the Apollo Metric Camera
images. This work was funded by the NASA Lunar Advanced Science and
Exploration Research (LASER) program grant \#07-LASER07-0148, NASA
Advanced Information Systems Research (AISR) program grant
\#06-AISRP06-0142, and by the NASA ESMD Lunar Mapping and Modeling
Program (LMMP). We would also like to thank James Burke and 
Bradley Bell at the University of Washington for guidance
in developing the optimization algorithm.

{\small
\bibliographystyle{ieee}
\bibliography{BAbib}
}

\end{document}